\documentclass[aps,prd,a4paper,10pt,superscriptaddress,preprintnumbers]{revtex4}

\usepackage{enumerate}
\usepackage{amsmath,amssymb,bm}
\usepackage[dvips]{graphicx}
\usepackage{braket}
\usepackage{url}
\usepackage{color}
\usepackage{mathrsfs} 
\usepackage{natbib}   

\usepackage{layout}
\allowdisplaybreaks[4]

\def\apss{Astrophys. Space Sci.}
\def\mnras{Mon. Not. R. Astron. Soc.}

\def\rot{{\rm rot}}
\def\gw{{\rm gw}}
\def\NS{{\rm NS}}
\def\omegarot{\omega_{\rm rot}}
\def\omegagw{\omega_{\rm gw}}

\begin{document}
\title{
A new estimation method for mass of an isolated neutron star using gravitational waves
}

\author{Kenji Ono}%
 \email{kenji@icrr.u-tokyo.ac.jp}
 \affiliation{
Department of Physics, 
Graduate School of Science, 
The University of Tokyo, Tokyo, 113-0033, Japan
}
\affiliation{ 
Institute for Cosmic Ray Research (ICRR), University of Tokyo, 5-1-5 Kashiwanoha, Kashiwa, Chiba 277-8582, Japan
}%

\author{Kazunari Eda}
\email{eda@resceu.s.u-tokyo.ac.jp}
\affiliation{
Department of Physics, 
Graduate School of Science, 
The University of Tokyo, Tokyo, 113-0033, Japan
}
\affiliation{
Research center for the early universe (RESCEU), 
Graduate School of Science, 
The University of Tokyo, Tokyo, 113-0033, Japan
}
\author{Yousuke Itoh}
 \email{yousuke_itoh@resceu.s.u-tokyo.ac.jp}
\affiliation{
Research center for the early universe (RESCEU), 
Graduate School of Science, 
The University of Tokyo, Tokyo, 113-0033, Japan
}

\begin{abstract} 
We investigate a possibility of estimating mass of an isolated rapidly 
rotating neutron star (NS) from a continuous gravitational wave (GW) signal 
emitted by the NS. When the GW passes through the gravitational potential 
of the NS, the GW takes a slightly longer time to travel to an observer 
than it does in the absence of the NS. Such a  time dilation effect holds 
also for photons and is often referred to as the gravitational time delay 
(or the Shapiro time delay). Correspondingly, the phase of the GW from the 
NS shifts due to the Coulomb type gravitational potential of the NS, and the 
resulting logarithmic phase shift depends on the mass, the spin frequency of the NS, 
and the distance to the NS. We show that the NS mass can, in principle, 
be obtained by making use of the phase shift difference between two modes of 
the continuous GW such as once and twice spin frequency modes induced by a freely 
precessing NS or a NS containing a pinned superfluid core. We estimate the 
measurement accuracy of the NS mass using Monte Carlo simulations and find that 
the mass of the NS with its ellipticity $10^{-6}$ at 1 kpc 
is typically measurable with an accuracy $20\%$ using Einstein Telescope. 
\end{abstract}

\preprint{RESCEU-2/15}
\maketitle

\section{Introduction}\label{Sec:intro}

Mass is a fundamental quantity of an astronomical object. In the case of a neutron 
star (NS), mass is important, among other aspects \cite{1999paln.conf.....W}, as it 
gives us a clue to study the equation of state of dense matter beyond the nuclear 
density, which is still uncertain through theoretical studies or experiments on the 
Earth (See, e.g., \cite{2012ARNPS..62..485L}).

Almost all the measurements of NS masses so far are limited for NSs in binaries. An 
exception is for the nearby isolated NS RX J 185635-754 \cite{1996Natur.379..233W},
where its radiation radius and gravitational redshift have been measured modulo 
detailed modeling of its atmosphere \cite{2002ApJ...564..981P} (See also the comments 
mentioned in the Sec. 4 of \cite{2011Ap&SS.336...67L} and Fig. 3 there).  Since an 
isolated star may have a different evolutionary history from a star in a binary 
(namely due to their mutual mass transfer), the former may follow a different mass 
function than the latter and hence it is interesting to measure masses of as many 
isolated NSs as possible. This paper proposes a new method to measure a mass of an 
isolated rapidly rotating NS that emits gravitational wave (GW).

We expect direct detection of GWs within the next decade \cite{2010CQGra..27q3001A}. 
The second-generation gravitational wave detectors such as KAGRA \cite{Aso:2013eba}, 
advanced Laser Interferometric Gravitationalwave Observatory (aLIGO) \cite{2010CQGra..27h4006H}, 
and advanced Virgo (aVirgo) \cite{2015CQGra..32b4001A} are under construction. 
Conceptual studies of third-generation gravitational wave telescopes are also on-going 
\cite{2010CQGra..27s4002P,2011CQGra..28i4013H}. One of the most important family of GW 
sources for those GW detectors is a compact binary coalescence (CBC). Detection of GW 
from a CBC event enables us to measure masses of compact stars in the binaries to 
sub-percent level.  A possibility is actively discussed where we can study internal
structure of the compact stars (if they are regular ones) through the tidal disruption 
of the binary stars during the binary merger (\cite{2011GReGr..43..409A} for a review). 
However, again, the estimates of masses possible from detection of GWs from CBC events are 
only for stars in binaries. We here focus on another type of GW sources, namely, isolated 
rapidly rotating non-axisymmetric NSs and show that we can estimate masses of isolated NSs 
by detection of those sources with one of proposed third-generation telescopes, namely, 
Einstein Telescope (ET) \cite{2010CQGra..27s4002P}.

In fact, our method proposed in this paper utilizes the fact that phase of GW from an isolated 
rapidly rotating compact star is modulated by the Coulomb type gravitational field of the star 
itself. This logarithmic phase shift is well-known in quantum mechanics \cite{messiah1958quantum}, 
and partly incorporated in the GW waveform from CBC \cite{Blanchet:1992br}. This Coulomb type phase
shift corresponds to the famous gravitational time delay (or the Shapiro time-delay when it is 
observed in a binary system).  But this phase shift seems neglected so far in the context of the 
studies searching for GWs from isolated pulsars. Then our idea to extract mass $M_{\NS}$ of an 
isolated pulsar at distance $r$ from GW observation is as follows. In the case of GW, the Coulomb 
phase shift take a form of $\delta \Phi(\omegagw) = 2\omegagw M_{\NS}\ln (2\omegagw r)$ for GW with 
frequency $\omegagw$ \cite{Asada:1997zu}.  If we detect two GW modes at frequencies $\omega_{\gw,1}$ 
and $\omega_{\gw,2}$  from the same pulsar, we could find the mass by $\delta \Phi(\omega_{\gw,1}) - 
K \delta \Phi(\omega_{\gw,2}) = 2\omega_{\gw,1} M_{\NS}\ln K$ where $K \equiv \omega_{\gw,1}/\omega_{\gw,2}$. 
Before examining the feasibility of this idea, we first mention whether a pulsar  could emit continuous 
GW at (more than) two frequencies.

A rotating tri-axial object with its rotational frequency $\omegarot$ emits GW mainly at frequency 
$2\omegarot$ when the rotational axis of the object is aligned to the one of the principal axis of 
its moment of inertia. If the former axis is not aligned to any one of the latters and this object 
rotates as a rigid body as a whole, this object will freely precess and emits GW at frequencies close
to $\omegarot$ and $2 \omegarot$ \cite{Zimmermann:1979ip}.  When the former axis is not aligned to 
any one of the latters and this object has a pinned superfluid layer(s), Jones proposed a possibility 
that this object could steadily rotate at a constant frequency and emits GW at frequencies close 
to $\omegarot$ and $2 \omegarot$ \cite{2010MNRAS.402.2503J}. The difference between the two scenarios 
\cite{Zimmermann:1979ip} and \cite{2010MNRAS.402.2503J} is that the latter does not freely precess and 
hence it shows no electromagnetic signature of free precession.

Observational evidences of existence of freely precessing NSs are recently  reported by Makishima {\it et al.} 
\cite{2014PhRvL.112q1102M,2015arXiv150107684M} where they found phase modulations in pulsations of the 
magnetar 4U 0142+61 and 1E 1547.0--5408 using the Suzaku X-ray observatory. They interpreted that those 
phase modulations are due to free precessions of those objects. There are several other objects that show 
signatures of free precessions, but it is not clear that current theoretical understandings of NS 
interior structures allow a free precession sustainable long enough so that we have any practical chance to 
observe it at all (See, e.g., \cite{2001MNRAS.324..811J,2002MNRAS.331..203J,2012MNRAS.420.2325J} for discussion 
and references therein). This in turn means that the findings by Makishima {\it et al.} and further detections 
of freely precessing NSs give us an insight onto the interior structure of NSs and drive our theoretical 
understandings on it.

Indeed, majority of pulsars do not show any clear signature of free precession. However, Jones pointed out 
that a NS may have a pinned superfluid. Jones then advocated a theoretical possibility that a NS could rotate 
steadily even when the axis of the moment of inertia of its solid outer crust is misaligned to its rotational 
axis  \cite{2010MNRAS.402.2503J}. As a result, such a NS emits GW at two frequencies close to $\omegarot$ and 
$2\omegarot$ without showing any electromagnetic signature of free precession. This proposal by Jones motivates 
Bejger and Kr\'olak \cite{2014CQGra..31j5011B} to study a search method for GWs from known pulsars at once and 
twice their spin frequencies. Motivated by these observations and the theoretical proposal by Jones, we here assume 
a pulsar that emits GW at frequencies $\omega_{\gw,1}\simeq \omegarot$ and $\omega_{\gw,2} = 2\omega_{\gw,1} \simeq 2\omegarot$ 
to study feasibility of our idea of estimating the pulsar mass using Coulomb phase shifts. The GW radiation at 
the frequency $\omega_{\gw,1} \simeq \omegarot$ is sometimes called a ``wobble mode'' or ``wobble radiation'' for a
freely precessing NS. Here we would like to include a non-precessing NS that emits GW at that frequency, we here 
call this mode ``the first harmonic mode'' while the GW at $2\omega$ ``the second harmonic mode'' (or ``the first overtone mode'').
Applications of our idea to GW modes other than the ones considered here or extensions to the case where multiple 
GW modes more than two are available should be straightforward.

The rest of this paper is organized as follows. In Sec. \ref{Sec:Formulation}, we construct a GW subject to a 
gravitational Coulomb phase shift due to the gravitational potential created by the GW source. In Sec.\ref{Sec:MassAccuracy}, 
we apply our formulation to an isolated NS and examine measurement accuracies of the NS parameters.  
Summary and conclusion are presented in Sec.\ref{Sec:Conclusion}. Throughout this paper, we adopt the geometric 
unit system in which both the light speed $c$ and the gravitational constant $G$ are unity.

\section{GW waveforms propagating through NS gravitational potential}\label{Sec:Formulation}
\subsection{Green's function}
In this section, we derive the explicit form of the gravitational phase shift for GW caused by the gravitational 
potential of the source itself. The derivation in this section is basically based on the section 2 in \cite{Asada:1997zu}. 
Consider an isolated GW source with mass $M$ and located at $r=0$.
The dominant static part of the exterior gravitational field far away from this object is well approximated by 
\begin{align}
 ds^2 &= - \left(1-\frac{2M}{r}\right)dt^2 + \left(1+\frac{2M}{r}\right)\delta_{ij}dx^idx^j.
\end{align}
A metric perturbation $\bar h^{\mu\nu} \equiv \eta^{\mu\nu}-\sqrt{-g}g^{\mu\nu}$ with respect to this metric 
satisfies the following wave equations 
\begin{align}
  \left[-\left(1+\frac{4M}{r}\right)\partial_t^2 + \boldsymbol{\nabla}^2 \right] \bar h^{\mu\nu} = 0
\end{align}
outside of the GW source in the harmonic gauge $\bar{h}^{\mu\nu}\mbox{}_{,\nu} = 0$ where $\eta^{\mu\nu}$ 
denotes the flat spacetime metric. Motivated by this observation and to capture the dominant effect of the 
gravitational field of the source onto the GW phase, we express the full Einstein equations including the 
source under the harmonic gauge condition in the following form, 
\begin{align}
 \Box_{\text{M}}\bar{h}^{\mu\nu}= - 16\pi \tau^{\mu\nu}. \label{Eq:WaveEq}
\end{align}
The effective energy-momentum pseudotensor $\tau^{\mu\nu}$ and the differential operator $\Box_{\text{M}}$ are defined as 
\begin{subequations}
\begin{align}
 &\Box_{\text{M}}  = \left[ - \left( 1 + \dfrac{4M}{r} \right) \partial_t^2 + \boldsymbol{\nabla}^2 \right], \\
 &\tau^{\mu\nu} = 
 \left(-g\right) \left( T^{\mu\nu} + t_{\text{LL}}^{\mu\nu} + t_{\text{M}}^{\mu\nu} \right), \\ 
 &t_{\text{M}}^{\mu\nu} 
 = -\dfrac{1}{16\pi} \left[ \bar{h}^{\mu\alpha}{}_{,\beta} \bar{h}^{\nu\beta}{}_{,\alpha} 
    + 2\bar{h}^{0j} \bar{h}^{\mu\nu}{}_{,0j} + \bar{h}^{jk} \bar{h}^{\mu\nu}{}_{,jk}
    + \left( \bar{h}^{00} - \dfrac{4M}{r} \right) \bar{h}^{\mu\nu}{}_{,00} \right],
\end{align}
\end{subequations} 
where $t_{\text{LL}}^{\mu\nu}$ denotes the Landau-Lifshitz pseudotensor \cite{landau1975classical} and $M$ 
denotes the total mass of the source. It should be noted that the term $\left(4M/r\right) \bar{h}^{\mu\nu}{}_{,00}$ 
appears in the both sides in Eq. (\ref{Eq:WaveEq}). Because the $00$ component of the metric perturbation is evaluated 
in the wave-zone as $\bar{h}^{00} = 4M/r$ to the lowest order in the weak field approximation, we intentionally added 
this term to the both sides in Eq. (\ref{Eq:WaveEq}). Introducing the ``potential barrier'' $4M/r$ in the wave operator, 
we can take into account the deviation of the curved spacetime light cone from the flat spacetime one to the lowest order 
in the weak field approximation. As will be seen in the end of this section, this mass term brings the gravitational 
phase shift in the GW phase. 

Equation. (\ref{Eq:WaveEq}) can be formally solved by using the retarded Green's function $G^{(+)}_{\text{M}} \left(x, x' \right)$ as,
\begin{align}
 \bar{h}^{\mu\nu} \left(x\right) 
 = -16\pi\int d^{4}x' \ G^{(+)}_{\text{M}} \left(x,x'\right) \tau^{\mu\nu}\left(x'\right), \label{Eq:h_intGtau}
\end{align}
where the retarded Green's function obeys the following equations. 
\begin{subequations}
\begin{align}
 &\Box_{\text{M}} G_{\text{M}}^{(+)} \left(x, x' \right) = \delta^{(4)} \left( x- x' \right), \label{Eq:G_diff}\\
 &G_{\text{M}}^{(+)} \left(x,x'\right)
 = \sum_{\ell m}\int d\omega \ e^{i\sigma_{\ell}} \text{sgn} \left(\omega \right) 
   \left[\Psi^{+\omega \ell m} \left(x\right) \Psi^{S\omega \ell m \ast}\left(x'\right)H \left(r-r'\right) 
   + \Psi^{S\omega \ell m}\left(x\right)\Psi^{+\omega \ell m \ast}\left(x'\right)H\left(r'-r\right)\right]. \label{Eq:G_sol}
\end{align}
\end{subequations} 
The functions $\Psi^{+\omega \ell m}$ and $\Psi^{S\omega \ell m}$ are composed of the spherical Coulomb wave functions 
$u^{(+)}\left(\rho\right)$ and $F_{\ell}\left(\rho\right)$ \cite{messiah1958quantum}:  
\begin{subequations}
\begin{align} 
 &\Psi^{+\omega \ell m}\left(x\right)=\sqrt{\frac{|\omega |}{2\pi}}
  e^{-i\omega t}\rho^{-1}u_{\ell}^{(+)} \left(\rho \right)Y_{\ell m} \left(\theta,\phi\right), \label{Eq:Psi_+}\\
 &\Psi^{S\omega \ell m}\left(x\right)=\sqrt{\frac{|\omega |}{2\pi}}
  e^{-i\omega t}\rho^{-1}F_{\ell}\left(\rho \right)Y_{\ell m} \left(\theta,\phi\right), \label{Eq:Psi_S}
\end{align}
\end{subequations}
where $\sigma_{\ell }= \mathrm{arg}\Gamma \left(\ell+1-2iM\omega\right)$, $\rho =  r\omega$, and $H\left(x\right)$ denotes 
the Heaviside step function. The radial functions $u_{\ell}^{(+)} \left(\rho\right)$ and $F_{\ell}\left(\rho\right)$ can be 
evaluated in the wave zone or the near zone as follows. 
\begin{subequations}
\begin{align}
 u^{(+)}_{\ell} \left(\rho\right)
  &\to \mathrm{exp} \left[ i\left(\rho+2M\omega\mathrm{ln}2\rho -\frac{1}{2}\ell \pi\right)\right] 
   \hspace{2em} \mathrm{for} \hspace{1em} \rho \rightarrow\infty, \label{Eq:ul_appx}\\
 F_{\ell} \left(\rho\right)
  & \to c_{\ell}\rho^{\ell+1} 
   \hspace{2em} \mathrm{for} \hspace{1em} \rho \rightarrow 0,\label{Eq:F_appx}
\end{align}
\end{subequations}
with  $c_{\ell} = 2^{\ell} e^{\pi M \omega} \left| \Gamma\left(\ell+1-2iM\omega\right)\right| / \left(2\ell+1\right) !.$
Since the distance to the source is much greater than the size of the source $r\gg r'$, 
substituting Eqs.(\ref{Eq:Psi_+})-(\ref{Eq:F_appx}) into Eq.(\ref{Eq:G_sol}) brings the Green's function to the following form, 
\begin{align}
 G^{(+)}_{\text{M}} \left(x,x'\right)
 &= \sum_{\ell} \dfrac{\left(-i\right)^{\ell} n^{\langle L \rangle} x'_{\langle L \rangle}}{2\left(2\pi \ell !\right)^2 r} 
    \int d\omega \ \Gamma\left(1+\ell-2iM\omega\right) e^{\pi M\omega} \omega^{\ell} e^{-i \left(t-r-2M\ln 2\omega r -t' \right)\omega }
    + \mathcal{O}\left(r^{-2}\right), \label{Eq:G+_finalform}
\end{align} 
where the angle bracket attached to the indices denotes the symmetric-trace-free operation, the suffix $L$ stands for $\ell$ 
tensorial products of vectors, and the unit vector $n^j$ is defined as $n^j = x^j / r$. Combining Eqs. (\ref{Eq:h_intGtau}) 
and (\ref{Eq:G+_finalform}), we obtain the GW waveforms affected by the gravitational potential of the source. The authors of 
Ref. \cite{Asada:1997zu} then adopted slow motion approximation and expanded this Green's function in $M\omega$ and arrived at 
the GW waveform including the tail term which was first derived by Blanchet and Damour \cite{Blanchet:1992br}. They then pointed out 
that the dominant part of the tail term originates from the deviation of the true light cone from the flat light cone due to the 
gravitational potential of the GW source itself, which appears as the phase shift $2iM\omega\ln 2\omega r + i\sigma_{\ell}$ 
in Eq. (\ref{Eq:G+_finalform}). 
Since we focus on this gravitational phase shift for GWs, we drop the $M \omega $ term in the amplitude  but retain the $M \omega$ 
term in the phase up to the first order in our analysis.

Finally we note that this phase shift corresponds to the well-known gravitational time delay (or the Shapiro time delay in the case 
of a binary). The GW propagates along the true light cone of the curved spacetime, and so takes a slightly longer time to travel to 
the detector than it does in the absence of the source.

\subsection{Continuous GWs through NS gravitational potential}
Let us now assume an isolated rapidly rotating NS with mass $M_{\text{NS}}$ as a GW source. As mentioned in the introduction, an isolated 
rapidly rotating NS would emit GWs at distinct frequencies at the same time. We define the volume integral of the stress-energy tensor 
$T^{jk}$ as, 
\begin{align} 
 S^{jk}\left(t\right) = \int d^3 x \ T^{jk}\left(t, \boldsymbol{x}\right). \label{Eq:Sjk_def}
\end{align}
If the source has different frequency modes $\omega_1, \omega_2, \cdots$, the Fourier component of $S^{jk}\left(t\right)$ can be decomposed 
into the sum of the delta function, 
\begin{align}
 \tilde{S}^{jk} \left(\omega \right) = \sum_{n} 2\pi \tilde{S}^{jk}_n
 \delta \left(\omega - \omega_n \right) \label{Eq:tildeSjk_def}, 
\end{align}
where $S^{jk}_n$ is the coefficient corresponding to the $n$-th mode. Combining Eqs. (\ref{Eq:h_intGtau}), 
(\ref{Eq:G+_finalform})-(\ref{Eq:tildeSjk_def}), we arrive at the GW waveform including the gravitational
phase shift in the quadrupole approximation, 
\begin{subequations} 
\begin{align}
 \bar{h}^{jk} \left(t\right)
 &= \dfrac{4}{r} \sum_{n} \tilde{S}^{jk}_n
 e^{-i\Psi_n\left(t\right)}, \label{Eq:GWwaveform} \\
 \Psi_{n} \left(t\right) 
 &= \left( t - r - 2M_{\text{NS}}  \left[\ln (2 r \omega_n)   + C\right] \right) \omega_n.  \label{Eq:GWphase}
\end{align}
\end{subequations} 
The constant  $C$ is the Euler's number. 

It is easy to see that when $M_{\text{NS}} \rightarrow 0$, Eq. (\ref{Eq:GWwaveform}) with 
Eq. (\ref{Eq:GWphase}) reduces to the well-known quadrupole gravitational waveform as expected. 
Inclusions of higher order mass multipoles or current multipoles, if necessary, are straightforward.

\section{Measurement accuracy of NS mass}\label{Sec:MassAccuracy}
In this section, we apply the GW waveform including the gravitational phase shift effect to an 
isolated rapidly rotating NS that emits GWs at two frequencies, $\omega_{\gw,1} = \omega$ and 
$\omega_{\gw,2} = 2\omega$.

\subsection{GWs from an isolated rapidly rotating non-axisymmetric NS}\label{Subsec:GWwaveform_freelyprecessingNS}
A non-axisymmetric NS rotating around its principal axis produces the quadrupole GW radiation with 
a frequency $2\omega_{\text{rot}}$, where $\omega_{\text{rot}}$ denotes the rotational frequency of 
the NS. However, the angular momentum axis of the NS does not generally align with its principal axis 
because of, for example, the strong toroidal magnetic field inside the NS \cite{2002PhRvD..66h4025C}. 
When the NS deviates from the spherical symmetry, the misalignment between the angular momentum axis 
and the principal axis in general leads to  a free precession of the NS. It would also be possible that 
the NS steadily rotates at $\omega_{\rot}$ at the fixed rotational axis in an inertial frame if it has 
a pinned superfluid component inside it \cite{2010MNRAS.402.2503J}.
Those two kinds of NSs radiate the GW with two frequency modes $\omega_{\gw,1} = \omega$ and 
$\omega_{\gw,2} = 2\omega$ which we call the first harmonic mode and the second harmonic mode, 
respectively \cite{Zimmermann:1979ip,2010MNRAS.402.2503J}.

The GW waveforms in these two scenarios are generally different from each other for tri-axial NSs 
\cite{Zimmermann:1979ip,Maggiore:2007,2010MNRAS.402.2503J,2014CQGra..31j5011B,2015arXiv150105832J}. 
. Interestingly, in the case of bi-axial NSs, those two families of the GW waveforms take the same 
form (shown e.g., in \cite{Jaranowski:1998qm}), which we adopt in this paper to apply our result 
to both scenarios.

We then adopt the following GW waveforms
\begin{subequations}
\begin{align}
 h_{+}      &= A_{+,1}\cos\Psi_{1} + A_{+,2} \cos\Psi_{2},          \label{Eq:GWwaveform_mode1}\\
 h_{\times} &= A_{\times,1}\sin\Psi_{1} + A_{\times,2} \sin\Psi_{2} \label{Eq:GWwaveform_mode2}
\end{align}
\end{subequations}
where $A$ and $\Psi$ denote the amplitude and the phase and are defined as 
\begin{subequations}
\begin{align}
 A_{+,1}     &= \dfrac{1}{4} h_{0}\sin 2\theta \sin\iota\cos\iota,               \label{Eq:GWamplitude_p1}\\
 A_{+,2}     &= \dfrac{1}{2} h_{0}\sin^{2}\theta \left( 1+\cos^{2}\iota \right), \label{Eq:GWamplitude_p2}\\
 A_{\times,1}&= \dfrac{1}{4} h_{0} \sin 2\theta \sin\iota,                       \label{Eq:GWamplitude_c1} \\
 A_{\times,2}&= h_{0}\sin^{2}\theta\cos\iota,                                    \label{Eq:GWamplitude_c2}\\
 \Psi_1 &=  \left( t-r-2M_{\text{NS}} \ln 2r\omega \right) \omega + \varphi_0,      \label{Eq:Psi_rot_1}\\
 \Psi_2 &= 2\left( t-r-2M_{\text{NS}} \ln 4r\omega \right) \omega + 2\varphi_0.      \label{Eq:Psi_rot_2}
\end{align}
\end{subequations} 
The angle $\iota$ is called the inclination defined as the angle between the rotational axis and 
the line-of-sight, and $\theta$ is the misalignment angle between the principal axis and the angular momentum 
axis. The angle $\varphi_0$ is the GW phase at the reference time $t=0$ at the NS \footnote{There is 
no Coulomb phase shift when the GW leaves the pulsar at $r =0$.}. The overall amplitude $h_0$ is 
given by $h_{0}=  4\varepsilon I \omega^2 /r$ where $I$ and $\varepsilon$ 
are the principle moment of inertia and the ellipticity, respectively. Note 
that the log-term appears in the GW phase due to the gravitational potential of the NS.

\subsection{How to extract the NS mass from GW signals}
In this paper, we take an example of GWs from an isolated rapidly rotating non-axisymmetric NS 
to demonstrate the measurement accuracy of the NS mass. A NS would emit GWs of several modes 
such as the first and the second harmonic modes mentioned in this paper, wobble mode if the spin 
axis of the NS precesses, $f$-mode and $r$-mode if the corresponding stellar oscillations are 
excited. If the initial GW phases of two modes at the NS are known, or the ratio of the two 
GW phases is equal to the ratio of the GW frequencies of the two mode, the NS mass can be estimated  
by subtracting the GW phase of one mode from the other with the ratio multiplied appropriately. In
the case of the first and second harmonic modes considered in this paper, this assumption is 
guaranteed because the both modes are generated by the deformation of the crust of the NS. Multiplying 
$\Psi_1$ by 2 Subtracting Eq. (\ref{Eq:Psi_rot_2}) from Eq. (\ref{Eq:Psi_rot_1}), we obtain 
\begin{align}
 2\Psi_{1} - \Psi_{2} = 4M_{\text{NS}} \omega \ln 2.  \label{Eq:diff_phase}
\end{align} 
Equation. (\ref{Eq:diff_phase}) indicates that the difference between the two modes separates the NS 
mass and the distance to the NS. Therefore, the NS mass alone can be estimated by GW observations.  
However, the accuracy of the constant phase is typically of the order of $0.1$ for a signal to noise 
ratio (SNR) 10 even in targeted searches for known pulsars where $\omegarot (\simeq \omega_{\gw,1})$ 
is well-known by electromagnetic observations. Then, the second generation GW detectors would have 
a little chance of detecting the log-term appeared in Eq. (\ref{Eq:diff_phase}) with enough sensitivity 
to estimate the NS mass. On the other hand, a third generation GW detector whose sensitivity is about 
an order of magnitude better than that of the second generation ones could determine the NS mass from 
the GW signal. 

As will be discussed  in the next section, we consider targeted searches for GWs from ``known'' rapidly 
rotating non-axisymmetric NSs by a third generation detector. Those objects are ``known'' in the sense 
that electromagnetic observations and/or the second generation detectors will have detected them before 
the search using third generation detectors. We then assume that the positions of the NSs in the sky and 
its GW frequencies (or rotation frequncies) and their times derivatives are known in advance. Those 
parameters are not search parameters in the following hypothetical search with a third generation GW detector.

\subsection{Monte Carlo simulations}
In this section, we investigate how accurately the mass of the NS can be determined by GW observations. 
Measurement errors are estimated by using the Fisher analysis which is briefly  summarized in Appendix 
\ref{Sec:FisherAnalysis}. A GW signal from an isolated rapidly rotating non-axisymmetric NS is given by  
Eqs. (\ref{Eq:GWsignal_decomposition})-(\ref{Eq:DopplerPhase}). The GW signal is characterized by 6 
waveform parameters $\boldsymbol{\lambda} = \left\{ h_0, \theta, \iota, \psi, \phi_0,  M_{\text{NS}} \right\}$ 
in targeted searches. 
To demonstrate the accuracies of the waveform parameters $\boldsymbol{\lambda}$, we perform Monte Carlo 
simulations for the observation time $T_{\text{obs}} = 3$ years in which we treat $\alpha,\delta,\theta,\iota$, 
and $\psi$ as random variables. Here, the sky position of the NS is specified by two parameters, the right 
ascension $\alpha$ and the declination $\delta$. The NS mass $M_{\text{NS}}$ is assumed to be 1.4 $M_{\odot}$. 
We adopted $\varepsilon = 10^{-6}$ \cite{2009PhRvL.102s1102H}, $I = 10^{38} \ {\rm kg}\cdot{\rm m}^2$, and selected 
distances and frequencies to compute the amplitude $h_0$. For each simulation, we generate 10000 sets of random 
variables each of which is distributed according to the uniform distribution.

\begin{figure}[htbp] 
\centering 
\includegraphics[width=16cm,clip]{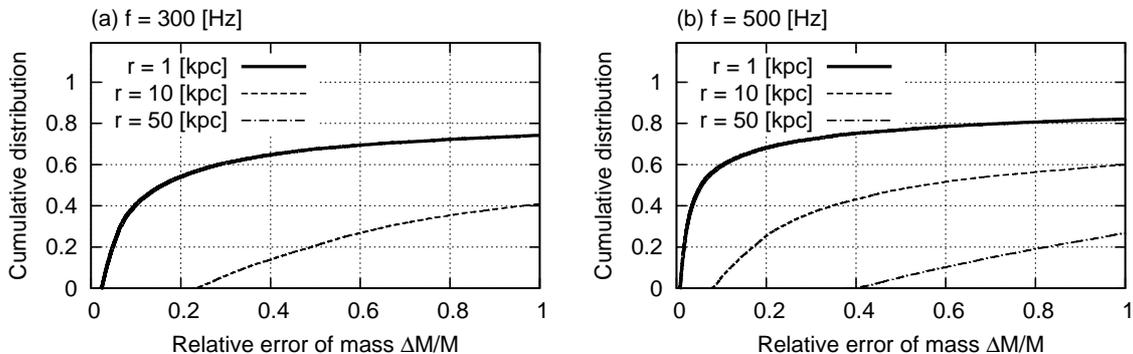}
\caption{\label{Fig:mass_accuracy}
The cumulative distribution function as a function of relative error of NS mass $\Delta M_{\text{NS}}/M_{\text{NS}}$ for 
two different GW frequencies (a) 300 Hz and (b) 500 Hz. The solid line, 
the dashed line, and the dashed-dotted line corresponds to NSs at the distances $r=1$ kpc, $10$ kpc, and $50$ kpc respectively. 
For NSs at  the distance of $r=50$ kpc that emit GW at the first harmonic GW mode frequency of $f_{\gw,1} = 300$ Hz, 
measurement accuracy becomes more than 100\%.
}
\end{figure}

Figure. \ref{Fig:mass_accuracy} displays the cumulative distribution functions of relative errors in the NS mass 
estimation $\Delta M_{\text{NS}}/M_{\text{NS}}$ from our Monte Carlo simulations for ET 
observations. We investigated the cumulative distribution functions for two different GW frequencies 300 Hz and 500 
Hz. These figures show, for example, in the case of the GW signal with $f_{\rot} \simeq f_{\gw,1} 
= \omega_{\gw,1}/(2\pi) = 500$ Hz from NSs at $r = 1$ kpc in ET observations, that masses of about a 70 percent of 
isolated rapidly rotating non-axisymmetric NSs are measurable with an accuracy of $\Delta M_{\text{NS}}/M_{\text{NS}} 
\simeq  0.2$. It can be detected with $\text{SNR} =  {\cal O}(10)$ by a single second generation detector because the sensitivity 
of ET is about an order of magnitude better than that of second generation one. Therefore, if an isolated rapidly 
rotating non-axisymmetric NS is detected by a network of second generation detectors, the mass of the NS can be determined 
with an accuracy of $\Delta M_{\text{NS}}/M_{\text{NS}} = \mathcal{O} \left(0.1\right)$ in a single third generation detector. 
Also, Fig.\ref{Fig:mass_accuracy} indicates that the accuracy of the NS mass becomes better and better as the GW frequency 
increases. This behavior can be traced to Eq. (\ref{Eq:diff_phase}). The effect of the gravitational phase shift is 
proportional to the rotational frequency. So the NS mass can be more easily extracted from the GW signal with the higher 
frequency.

\begin{figure}[htbp] 
\centering 
\includegraphics[width=16cm,clip]{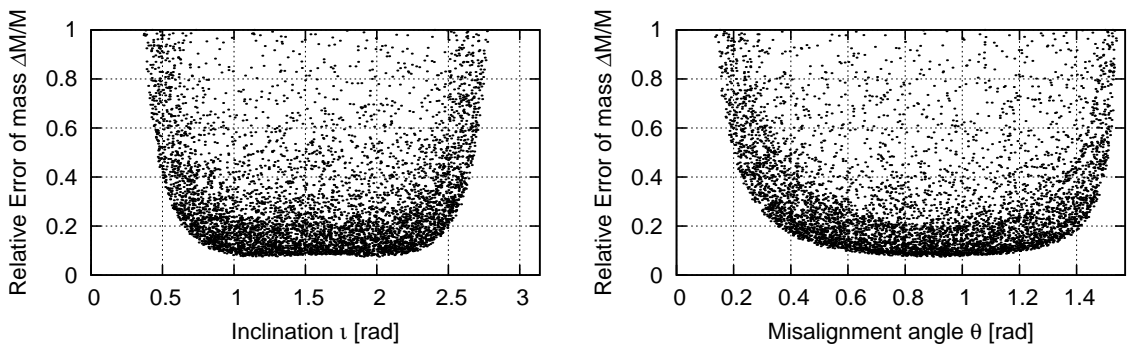}
\caption{\label{Fig:scatterplot}
The scatter-plot of relative error of NS mass $\Delta M_{\text{NS}}/M_{\text{NS}}$ and inclination $\iota$ (left), misalignment 
angle $\theta$ (right). The GW signals are assumed to have the GW frequency 500 Hz from NSs at $r = 10$ kpc. 
 } 
\end{figure}

Figure. \ref{Fig:scatterplot} shows that the scatter-plot of the NS mass accuracy $\Delta M_{\text{NS}}/M_{\text{NS}}$ 
as a function of the inclination $\iota$ (left) and the angle $\theta$ (right) for GW signals with $f_{\gw,1} = 500$ Hz from 
NSs at $r = 10$ kpc. The left panel of Fig. \ref{Fig:scatterplot} indicates that the NS mass cannot be estimated at all at 
$\iota = 0, \pi$ while well estimated at around $\iota=\pi/2$. This is because the amplitudes of the first harmonic mode in 
both the plus and cross polarizations vanish at $\iota=0,\pi$ from Eqs. (\ref{Eq:GWamplitude_p1}) and (\ref{Eq:GWamplitude_c1}). 
In these cases, since the GW signal has only the second harmonic mode, the effect of the gravitational phase shift is unobservable. 
The right panel in Fig. \ref{Fig:scatterplot} indicates that the NS mass cannot be determined for $\theta = 0, \pi/2$. 
This fact can be interpreted in the same way as the left panel of Fig. \ref{Fig:scatterplot}.


\section{Conclusion}\label{Sec:Conclusion}
In this paper, we have focused on a gravitational phase shift for a GW from an isolated rapidly rotating non-axisymmetric NS. When the 
GW passes through the gravitational potential created by the NS, the GW phase is stretched due to the gravitational phase shift effect 
depending on both its rotational frequency and its mass. We have constructed the explicit form of the resulting phase shift based on 
Ref. \cite{Asada:1997zu} and obtained Eq. (\ref{Eq:GWphase}).  We have shown that if the GW from the NS has different 
frequency modes such as the first and the second harmonic modes as proposed in \cite{Zimmermann:1979ip,2010MNRAS.402.2503J}, 
the mass of the isolated NS can be, in principle, determined by making use of the phase difference 
of the different modes. We have given an example of an isolated rapidly rotating non-axisymmetric NS to estimate the measurement accuracy 
of the NS mass using Fisher analysis in a targeted search. Performing Monte Calro simulations in 3-year ET observation, we have obtained 
a cumulative distribution function of a relative error of the NS mass $\Delta M_{\text{NS}}/M_{\text{NS}}$ for different GW frequencies 300
Hz and 500 Hz and different distances of $r = 1, 10$, and  $50$ kpc as shown in Fig. \ref{Fig:mass_accuracy}. The range of the NS mass can 
be typically delimited with an accuracy of $\Delta M_{\text{NS}}/M_{\text{NS}} = \mathcal{O} \left(0.1\right)$ with third generation detectors 
for such a GW signal that can be detected by a second generation GW detector with $\text{SNR} = {\cal O}(10$).

\begin{acknowledgments}
We thank Masatake Ohashi for helpful comments. This work is supported by Grant-in-Aid for Japan Society for the Promotion of Science, 
the JSPS Fellows Grant No. 26.8636 (K. E.), the JSPS Grant-in-Aid for Young Scientists Grant No. 25800126 (Y. I.), and the Ministry of Education, 
Culture, Sports, Science and Technology (MEXT) Grant-in-Aid for Scientific Research on Innovative Areas ``New Developments in Astrophysics
Through Multi-Messenger Observations of Gravitational Wave Sources'' (Grant Number 24103005) (Y. I.).
\end{acknowledgments}

\appendix
\section{GW signals}
The GW waveforms from an isolated rapidly rotating non-axisymmetric 
NS are presented in Sec. \ref{Subsec:GWwaveform_freelyprecessingNS}. 
As can be seen in Eq. (\ref{Eq:GWwaveform_mode1}) and (\ref{Eq:GWwaveform_mode2}), 
the GW signals consist of two components with frequency $f_{\gw,1}$ and $2f_{\gw,1}$. 
It is convenient for our purpose to decompose the GW signals as follows \cite{Jaranowski:1998qm}. 
\begin{align}
 h\left(t\right) = \sum_{j=1}^{2}\sum_{k=1}^{4}A_{jk} h_{jk}\left(t\right). \label{Eq:GWsignal_decomposition}
\end{align}
 The amplitudes $A_{jk}$ are defined as 
\begin{subequations} 
\begin{align}
 &A_{11} = h_0 \sin 2\theta \left[ \dfrac{1}{8} \sin 2\iota \cos 2\psi \cos\phi_0 - \dfrac{1}{4} \sin\iota \sin 2\psi \sin \phi_0 \right], \\
 &A_{12} = h_0 \sin 2\theta \left[ \dfrac{1}{4}\sin \iota \cos 2\psi \sin \phi_0 + \dfrac{1}{8} \sin 2\iota \sin 2\psi \cos\phi_0  \right], \\
 &A_{13} = h_0 \sin 2\theta \left[  -\dfrac{1}{8}\sin 2\iota \cos 2\psi \sin \phi_0 - \dfrac{1}{4}\sin \iota \sin 2\psi \cos \phi_0  \right], \\
 &A_{14} = h_0 \sin 2\theta \left[ \dfrac{1}{4}\sin \iota \cos 2\psi \cos \phi_0 - \dfrac{1}{8} \sin 2\iota \sin 2\psi \sin \phi_0   \right], \\
 &A_{21} = h_0 \sin^2\theta \left[ \dfrac{1}{2}\left( 1 + \cos^2 \iota \right) \cos 2\psi \cos \left( 2\phi_0 + \phi_s \right) - \cos \iota \sin 2\psi \sin \left( 2\phi_0 + \phi_s \right)  \right], \\
 &A_{22} = h_0 \sin^2\theta \left[ \dfrac{1}{2}\left( 1 + \cos^2 \iota \right) \sin 2\psi \cos \left( 2\phi_0 + \phi_s \right) +  \cos \iota \cos 2\psi \sin \left( 2\phi_0 + \phi_s \right)   \right], \\
 &A_{23} = h_0 \sin^2 \theta \left[ - \dfrac{1}{2}\left( 1 + \cos^2 \iota \right) \cos 2\psi \sin \left( 2\phi_0 + \phi_s \right) -  \cos \iota \sin 2\psi \cos \left( 2\phi_0 + \phi_s \right)  \right], \\
 &A_{24} = h_0 \sin^2 \theta \left[ - \dfrac{1}{2}\left( 1 + \cos^2 \iota \right) \sin 2\psi \sin \left( 2\phi_0 + \phi_s \right) +  \cos \iota \cos 2\psi \cos \left( 2\phi_0 + \phi_s \right) \right]
\end{align}
\end{subequations}
where $h_0, \iota, \theta, \psi$ and $\phi_0$ are the overall amplitude, the inclination, 
the polarization phase, and the initial phase, respectively. 
The phase $\phi_s$ expresses the gravitational phase shift 
due to the gravitational potential of the NS and 
corresponds to Eq. (\ref{Eq:diff_phase}). It is described by 
\begin{align}
 \phi_s = 8\pi M_{\text{NS}} f_{\gw,1} \ln 2. 
\end{align}
The time-dependent parts $h_{jk} \left(t\right)$ in Eq.(\ref{Eq:GWsignal_decomposition}) 
characterize the shape of the GW signal and are defined as 
\begin{subequations}
\begin{align}
 &h_{j1}\left(t\right) = a\left( t\right) \cos \left[ j\Phi\left(t\right) \right], \\
 &h_{j2}\left(t\right) = b\left( t\right) \cos \left[ j\Phi\left(t\right) \right], \\
 &h_{j3}\left(t\right) = a\left( t\right) \sin \left[ j\Phi\left(t\right) \right], \\
 &h_{j4}\left(t\right) = b\left( t\right) \sin \left[ j\Phi\left(t\right) \right]
\end{align}
\end{subequations}
where the function $a\left(t\right)$ and $b\left(t\right)$ are modulation amplitudes given in \cite{Jaranowski:1998qm}. 
The GW phase after the barycentric corrections is expressed by 
\begin{align}
 \Phi\left(t\right) = 2\pi f_{\gw,1} \left( t + \boldsymbol{n} \cdot \boldsymbol{r}\left(t\right) \right) 
 \label{Eq:DopplerPhase}
\end{align}
where $\boldsymbol{n}$ denotes a unit vector pointing toward the NS from 
the solar system barycenter (SSB), and $\boldsymbol{r} \left(t\right)$ 
denotes the separation vector pointing from the detector to the SSB. 
Note that we neglected the spin-down effect in the GW frequency for simplicity. 
The decomposition of the GW signals in Eq.(\ref{Eq:GWsignal_decomposition}) 
separates known parameters and unknown parameters in targeted searches.

\section{Fisher analysis}\label{Sec:FisherAnalysis}
In this appendix, we give a brief review of the Fisher analysis \cite{Finn:1992wt,Cutler:1994ys}. 
The detector output $s\left(t\right)$ can be expressed by linear sum of 
the GW signal $h\left(t\right)$ and the detector noise $n\left(t\right)$, 
$s\left(t\right) = h\left(t\right) + n\left(t\right)$. 
When the noise is stationary and obeys the Gaussian distribution, 
measurement errors of the waveform parameters $\boldsymbol{\lambda}$ are found by 
\begin{align}
 \left( \Delta \lambda^j \right)_{\text{rms}}
 \equiv \sqrt{ \langle \left(\Delta \lambda^j\right)^2 \rangle }
 = \sqrt{ \left( \Gamma^{-1} \right)_{jj} }
\end{align}
for large signal-to-noise ratio where $\left( \Delta \lambda^j \right)_{\text{rms}}$
is the root-mean square of the waveform parameters $\Delta \lambda^j$. 
The matrix $\Gamma_{jk}$ is called the Fisher information matrix and is defined by 
\begin{align} 
 \Gamma_{jk} = \left( \partial_j h \Big| \partial_k h  \right) \label{Eq:DefFisher}
\end{align}
where $\left( \cdot | \cdot \right)$ denotes the noise-weighted inner product, 
and $\partial_j h$ denotes the derivative of the GW signal with respect to the waveform parameter $\lambda_j$. 
For instance, the inner product between two time functions $x \left(t\right)$ and $y\left(t\right)$ 
is expressed by 
\begin{align}
 \left( x | y \right) =  4 \text{Re} \int_{-\infty}^{\infty} \dfrac{\tilde{x} \left(f\right) \tilde{y}^{\ast} \left(f\right) }{S_n \left(f\right)} df. 
 \label{Eq:InnerProduct}
\end{align}
where $S_n(f)$ is the one-sided powerspectral density of the
detector. In this paper, we have adopted $S_n(f)$ of ET given
in the Table 1 of \cite{2009LRR....12....2S}. 

When the frequency of the signal is nearly constant, 
it is convenient to introduce a new inner product 
\begin{align}
 \left( x || y \right) \equiv
 \dfrac{2}{T_{\text{obs}}} \int_{-T_{\text{obs}}/2}^{T_{\text{obs}}/2} x\left(t\right) y \left(t\right) dt \label{Eq:NewInnerProduct}
\end{align}
where $T_{\text{obs}}$ denotes the observation time \cite{Jaranowski:1998qm}. 
Since the time scales of change in modulation amplitudes $a\left(t\right)$ and $b\left(t\right)$ are
of the order of the time scale of the Earth's rotation, 
the GW phase $\Phi \left(t\right)$ oscillates much more rapidly than modulation amplitudes. 
Then, the inner product between the amplitudes $h_{jk}$ can be well-approximated as
\begin{subequations}
\begin{align}
  & \left( h_{j1} || h_{j1} \right) \simeq \left( h_{j3} || h_{j3} \right) \simeq \dfrac{1}{2} \left( a || a \right), \label{Eq:Inner_aa} \\
  & \left( h_{j2} || h_{j2} \right) \simeq \left( h_{j4} || h_{j4} \right) \simeq \dfrac{1}{2} \left( b || b \right), \label{Eq:Inner_bb} \\
  & \left( h_{j1} || h_{j2} \right) \simeq \left( h_{j3} || h_{j4} \right) \simeq \dfrac{1}{2} \left( a || b \right). \label{Eq:Inner_ab}
\end{align}
\end{subequations}
Combining Eqs. (\ref{Eq:DefFisher})-(\ref{Eq:Inner_ab}), we obtain the Fisher matrix in our analysis, 
\begin{align}
 \Gamma_{jk} 
 &= \sum_{\ell =1,2} \dfrac{T_{\text{obs}}}{2S_n \left(\ell f_{\gw,1} \right)}
     \Big[ \left( \partial_j A_{\ell 1} \partial_k A_{\ell 1} + \partial_j A_{\ell 3 } \partial_k A_{\ell 3} \right) \left( a || a \right) 
     + \left( \partial_j A_{\ell 2} \partial_k A_{\ell 2} + \partial_j A_{\ell 4} \partial_k A_{\ell 4} \right) \left( b || b \right) \nonumber \\
 &\hspace{1em }
     + \left( \partial_j A_{\ell 1} \partial_k A_{\ell 2} + \partial_j A_{\ell 2} \partial_k A_{\ell 1} 
     + \partial_j A_{\ell 3} \partial_k A_{\ell 4} + \partial_j A_{\ell 4} \partial_k A_{\ell 3}  \right) \left( a || b \right) \Big]. 
\end{align}

\end{document}